\documentclass[a4paper]{jpconf}
\usepackage{graphicx}
\begin{document}
\title{Coherent manipulation of nuclear spins in the breakdown regime of
integer quantum Hall states}

\author{M. Kawamura$^1$, H. Takahashi$^1$, Y. Hashimoto$^2$, \\
S. Katsumoto$^{2,3}$,  and T. Machida$^{1,3}$}

\address{$^1$ Institute of Industrial Science, University of Tokyo,
Tokyo, Japan}
\address{$^2$ Institute for Solid State Physics,
University of Tokyo, Tokyo, Japan}
\address{$^3$ Institute for Nano Quantum Information Electronics, University of Tokyo,
Tokyo, Japan}

\ead{minoru@iis.u-tokyo.ac.jp}
\ead{tmachida@iis.u-tokyo.ac.jp}

\begin{abstract}
We demonstrate a new method for electrical manipulation of nuclear spins
utilizing dynamic nuclear polarization induced by quantum Hall effect breakdown. 
Nuclear spins are polarized and detected through 
the hyperfine interaction between a nuclear spin system
and a two-dimensional electron system located at an interface 
of GaAs/AlGaAs single heterostructure.
Coherent oscillations between the nuclear-spin quantum states are
observed by measuring the longitudinal voltage of the conductor.
\end{abstract}

Manipulation and detection of nuclear spins have attracted much interests
in research field of low-dimensional semiconductor systems
ever since the proposed use of nuclear spins as solid-state quantum bits
\cite{Kane1998}.
Dynamic nuclear polarization (DNP) is one of the essential techniques
to initialize nuclear-spin qubits.
So far, coherent manipulation of solid-state nuclear spins has been
demonstrated in several low-dimensional semiconductor systems
using different kinds of DNP techniques\cite{Machida2003, Yusa2005, Sanada2006}.
They utilize spin-flip process of electrons associated with
electron scatterings between spin-resolved quantum Hall 
edge channels\cite{Machida2003},
inter-domain scatterings between spin-polarized and -unpolarized domain structures
in fractional quantum Hall systems\cite{Yusa2005},
 and spin-selective excitation of electrons by circularly polarized laser beam
irradiation\cite{Sanada2006}.
In these DNP techniques, nuclear spins are polarized by the flip-flop process
of electron spins ${\boldmath S}$  and nuclear spins ${\boldmath I}$ 
in the hyperfine interaction,
$
{\cal H}_{\rm hyp} = A{\boldmath I}\cdot{\boldmath S}
 = A(I^{+}S^{-} + I^{-}S^{+})/2 + AI_{z}S_{z}
$,
where $A$ is the hyperfine constant.

In this paper, we report our novel method for coherent manipulation of 
nuclear spins.
We initialize nuclear spins
located at the interface of GaAs/AlGaAs heterostructure
by a DNP technique using breakdown of integer quantum Hall effect (QHE)\cite{Kawamura2007}.
Then, the nuclear-spin states are manipulated by 
irradiation of radio-frequency (rf) pulse magnetic field.
The manipulated nuclear-spin states are read out by standard dc voltage measurements
of the quantum Hall conductor.
By changing the rf-pulse width, we observe coherent evolution of the nuclear-spin states,
{\it i.e.} the Rabi oscillation.

\begin{figure}
	\begin{center}
		\includegraphics[width=8.5cm]{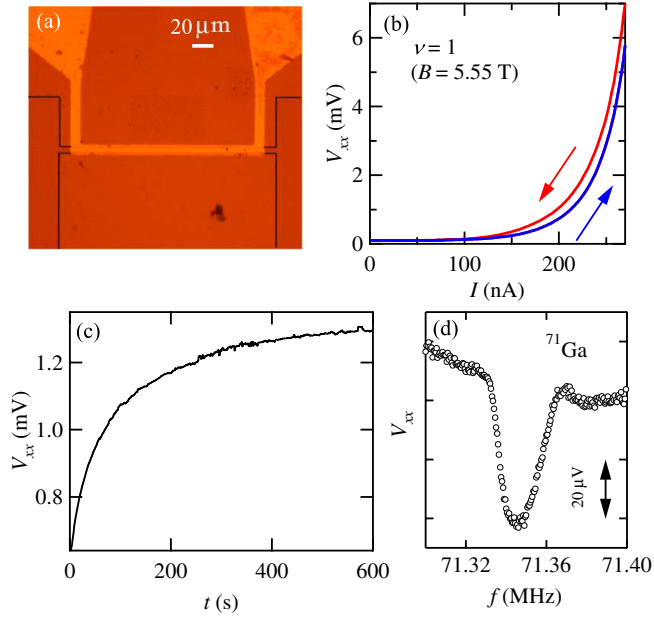}
	\end{center}
	\caption{\label{fig1}
		(a) Microphotograph of the Hall-bar device used in the present study.
		(b) Voltage-current characteristic curves at $\nu$ = 1 ($B$ = 5.55~T)
			obtained by sweeping $I$ in positive and negative directions.
		(c) Evolution of $V_{xx}$ after switching $I$ from 0~nA to 200~nA.
		(d) NMR spectrum of $^{71}$Ga obtained by measuring $V_{xx}$ at $I$ = 200~nA.
	}
\end{figure}

We study a Hall-bar device fabricated from a wafer
of GaAs/AlGaAs single heterostructue
with two-dimensional electron gas (2DEG) at the  interface.
The  electron mobility and the carrier density are 220~m$^2$/Vs and 
1.6~$\times$~10$^{15}$~m$^{-2}$, respectively.
Figure~\ref{fig1}(a) shows an optical micrograph of the Hall-bar device.
The central part of the Hall bar is covered with 
a 10-$\mu$m-wide Ti/Au metal strip.
The metal strip is used to apply rf-magnetic fields $B_{\rm rf}$ to the 2DEG.
All the measurements were performed using a standard dc four-terminal method
in a $^3$He-$^4$He dilution refrigerator at about 50~mK.

In order to initialize the nuclear-spin state, 
we use a DNP technique utilizing breakdown
of quantum Hall effect developed in our earlier study\cite{Kawamura2007}.
The QHE breakdown is a phenomenon
in which a dissipationless quantum Hall state
is broken by applying a bias current above a critical value $I_{\rm c}$.
When the quantum Hall state is broken,
electrons condensed in the Landau levels are excited
to the upper Landau level, giving rise to an abrupt increase in 
longitudinal voltage $V_{xx}$.
In a case of odd-integer QHE,
the electron excitations involve up-to-down spin flips.
The up-to-down flips of electron spins cause to 
flop nuclear spins upward through the hyperfine interaction.
The polarized nuclear spins reduce the spin-splitting energy
$\Delta E = |{\rm g}|\mu_{\rm B}B - A\langle I_z \rangle$,
where g is the g-factor for electrons (=$-0.44$ in GaAs) and
$\mu_{\rm B}$ is the Bohr magneton.
Hence, this process accelerates the QHE breakdown, leading to 
an increase in $V_{xx}$ and hysteretic $V_{xx}$-$I$ characteristic curves\cite{Kawamura2007}.
As demonstrated in our earlier experiment\cite{Kawamura2008},
nuclear spins in the bulk part of the quantum Hall conductor are polarized
because the current mainly flows in the bulk part of the conductor.

Figure~\ref{fig1}(b) shows $V_{xx}$-$I$ curves
at $B$ = 5.55~T ($\nu$ = 1) obtained by sweeping $I$ 
at a rate of 17~nA/s in positive
and negative directions. 
The down-sweep curve appears on the small current side of the up-sweep curve.
Clear hysteresis is observed above $I$ = 100~nA.
When $I$ is switched from 0~nA to 200~nA,
the value of $V_{xx}$ increases gradually over 600~s 
as shown in Fig.~\ref{fig1}(c).
By applying continuous-wave rf-magnetic field,
we obtain a nuclear magnetic resonance (NMR)  spectrum with a dip structure
at a frequency of 71.346~MHz [Fig.~\ref{fig1}(d)], 
which is the NMR frequency for $^{71}$Ga at this magnetic field .
NMR spectra for $^{69}$Ga and $^{75}$As were also obtained.
These results are consistent with our earlier experiments\cite{Kawamura2007}.
Since the amplitude of the in-plane component of $B_{\rm rf}$ 
(perpendicular to $B$)
is strong only in the region beneath the metal strip, nuclear spins
located under the metal strip are relevant to the NMR spectrum.

\begin{figure}
	\begin{center}
		\includegraphics[width=10cm]{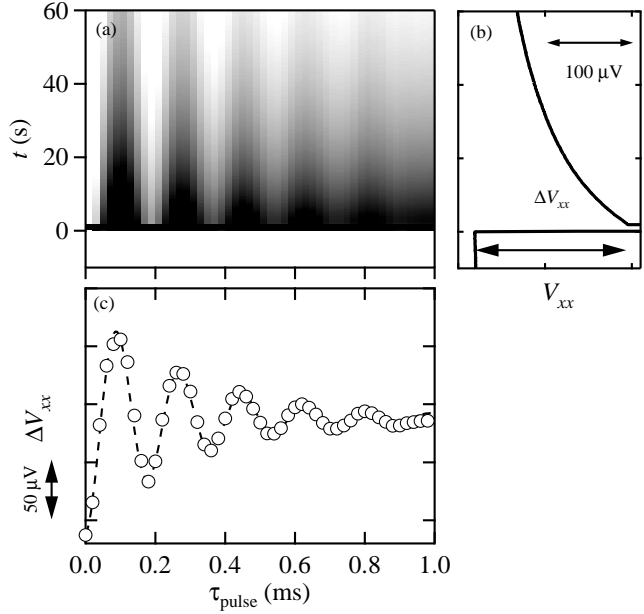}
	\end{center}
	\caption{\label{fig2}
		(a) Gray-scale plot of $V_{xx}$ as functions of rf-pulse duration $\tau_{\rm pulse}$
		and measurement time $t$.
		(b) A representative time evolution of $V_{xx}$ for $\tau_{\rm pulse}$ = 80~$\mu$s.
		The rf pulse is applied at $t$ = 0~s.
		(c) Rabi oscillation for $^{71}$Ga traced by the amplitude of $\Delta V_{xx}$.
		}
\end{figure}

Combining the above DNP technique with the pulse NMR technique,
we manipulate $^{71}$Ga nuclei coherently:
First, nuclear spins are polarized upward using the above-described
DNP technique utilizing the QHE breakdown.
Then, rf-magnetic field with pulse duration $\tau_{\rm pulse}$
(0 $\le$ $\tau_{\rm pulse}$ $\le$ 1 ms) are applied at $t$ = 0~s.
A representative time evolution of $V_{xx}$
is shown in Fig.~\ref{fig2}(b) for $\tau_{\rm pulse}$ = 80~$\mu$s.
The value of $V_{xx}$ decreases abruptly after the irradiation of
a rf pulse and recovers to the initial value in a time scale of several tens of  seconds.
The voltage drop after the rf-pulse irradiation (at $t$ = 5~s) 
is denoted as $\Delta V_{xx}$ as shown in Fig.~\ref{fig2}(b).

By repeating the above procedure with different $\tau_{\rm pulse}$,
we obtain a gray-scale plot of 
$V_{xx}$ as functions of $\tau_{\rm pulse}$ and measurement 
time $t$ as shown in Fig.~\ref{fig2}(a).
The value of $\Delta V_{xx}$ oscillates periodically as a function of $\tau_{\rm pulse}$
as plotted in Fig.~\ref{fig2}(c).
This oscillation corresponds to the coherent evolution 
of nuclear-spin states, {\it i.e.} the Rabi oscillation.
The amplitude of the Rabi oscillation in the present work
is more than 10 times larger than that in our previous experiment\cite{Takahashi2007},
in which a Hall-bar device with shorter channel length was used.
The amplitudes of the oscillations were almost the same
even at higher temperatures up to 1.5 K.

The oscillation decays gradually with a time scale of about 300~$\mu$s,
which is rather short for a coherence time of a nuclear spin system in GaAs.
The short decay time is probably attributed to the rather broad NMR spectrum (30 kHz in FWHM). 
Since $^{71}$Ga nucleus has a spin of $I$ = 3/2, 
we suspect that the electric quadrupole interaction causes the broad spectrum.

%
%
%

\ack
This work is supported by a Grant-in-Aid from MEXT,
the Sumitomo Foundation,
and the Special Coordination Funds
for Promoting Science and Technology.


\end{document}